\documentclass[12pt,preprint]{aastex}

\usepackage{psfig}
\usepackage{graphicx}
\usepackage{natbib}
\usepackage[english]{babel}  

\shorttitle{X-ray emitting ejecta in the Vela SNR}
\shortauthors{Miceli et al.}

\begin{document}

\title{Physical and chemical inhomogeneities inside the Vela SNR shell. Indications of ejecta shrapnels}

\author{M. Miceli\altaffilmark{1,2}, F. Bocchino\altaffilmark{1,2}, F. Reale\altaffilmark{1,2,3}}

\altaffiltext{1}{INAF - Osservatorio Astronomico di Palermo, Piazza del Parlamento 1, 90134 Palermo, Italy; miceli@astropa.unipa.it}
\altaffiltext{2}{Consorzio COMETA, Via S. Sofia 64, 95123 Catania, Italy}
\altaffiltext{3}{Dipartimento di Scienze Fisiche ed Astronomiche, Sezione di Astronomia, Universit\`a di Palermo, Piazza del Parlamento 1, 90134 Palermo, Italy}

\begin{abstract}
We present the results of the combined analysis of three XMM-Newton EPIC observations of the northern rim of the Vela SNR.
The three pointings cover an area of $\ga$ 10 pc$^2$ (at 250 pc) behind the main shock front and we aim at studying with high resolution the spatial distribution of the physical and chemical properties of the X-ray emitting plasma on this large scale.
We produce count-rate images and equivalent width maps of the Ne IX and Mg XI emission blends. We also perform a spatially resolved spectral analysis of a set of physically homogeneous regions.
We reveal physical and chemical inhomogeneities in the X-ray emitting plasma. In particular, we find large variations of the O, Ne, Mg, and Fe abundances. In some bright knots we also find unexpectedly enhanced Ne and Mg abundances, with values significantly larger than solar.
Our results support a possible association of a few X-ray emitting knots with previously undetected residuals of stellar fragments (i. e. shrapnels) observed, in projection, inside the Vela shell.
\end{abstract}


\maketitle

\section{Introduction}
\label{Introduction}

The study of middle-aged supernova remnants (SNRs) in the X-ray band is traditionally considered a powerful diagnostic tool to obtain information about the dynamics of shock-cloud interaction and the relevant properties of the interstellar and circumstellar medium. However, although the bulk of the X-ray emission in evolved remnants is associated with the post-shock interstellar medium (ISM), there is emerging evidence for the presence of X-ray emitting ejecta even in middle-aged SNRs. For example, there are claims for the presence of chemical inhomogeneities in the Cygnus Loop, observed in X-rays both at large (\citealt{mt99}, but see \citealt{lgw02}) and small (\citealt{lea04}) spatial scales. The metal abundances found in the Cygnus Loop are always lower than (or consistent with) the solar values, but the presence of such chemical inhomogeneities suggests that the X-ray emission cannot be completely associated with the shocked ISM.

In the Vela SNR there are indications of X-ray emitting ejecta. \citet{aet95} identified 6 ``shrapnels'' (labelled Shrapnel A-F), that are X-ray emitting features (with a characteristic boomerang shape) protruding beyond the primary blast wave. Shrapnel A and Shrapnel D have been recently observed with \emph{XMM-Newton}. \citet{kt05} found high O, Ne, and Mg abundances in Shrapnel D, thus confirming its association with a fragment of supernova ejecta. As for Shrapnel A, a significant Si overabundance has been observed, while O, Ne, Mg, and Fe have solar or subsolar abundances. These results seem to indicate that Shrapnel A is somehow different from Shrapnel D and \citet{kt06} conclude that the ejecta in Shrapnel A are strongly mixed with the swept-up ISM. 

The results obtained in the Cygnus Loop and in the Vela SNR open up the possibility to study the products of SN nucleosynthesis in large and near remnants.
Because of its proximity ($\sim 250$ pc, \citealt{bms99}, \citealt{csd99}), Vela SNR is an ideal target for this kind of studies, since it is possible to resolve fine structures at high spatial resolution. The age of this SNR has been estimated to be $\sim 10^{4}$ yr, in good agreement with the characteristic age of the pulsar PSR B0833-45 ($\sim 11200$ yr, \citealt{tml93}), which is associated with the remnant, as shown by \citet{wp80} and \citet{ws88}.

The northern rim of the Vela shell presents a patchy X-ray emission indicating a complex interaction between the blast wave shock and several ambient inhomogeneities. In \citet{mbm05} and \citet{mro06} (hereafter Paper I and Paper II) we studied a small isolated X-ray knot (Vela $FilD$) located near the northern shock front. 
The spectral analysis shows that two thermal components ($T_{I}\sim 10^{6}$ K, $T_{II}\sim 3\times 10^{6}$ K) of an optically-thin plasma in collisional ionization equilibrium (CIE) are present (Paper I). Detailed hydrodynamic modeling (Paper II) shows that in the $FilD$ knot the cooler component (the cloud ``core'') originates in the cloud material heated by the transmitted shock front. The hotter component (the cloud ``corona'') is the result of thermal conduction between the cloud and the hotter shocked intercloud medium.
Moreover, the peculiar orientation of the optical filament associated with the $FilD$ X-ray knot is naturally explained by our model as a result of thermal instabilities.

An open issue is related to the study of the metal abundances behind the shock front. As shown in Paper I, the abundances found in the $FilD$ region are unusual and not expected according to metal depletion and grain destruction behind shock wave models, showing Ne slightly overabundant, Fe underabundant, and solar O. It is then interesting to study how these abundances are distributed at different distances from the main shock front and how they are related to the abundances observed in the Vela shrapnels. There is, in fact, the concrete possibility that additional shrapnels are observed in projection inside the shell.

\begin{figure}[h!]
\centerline{\includegraphics[width=12 cm]{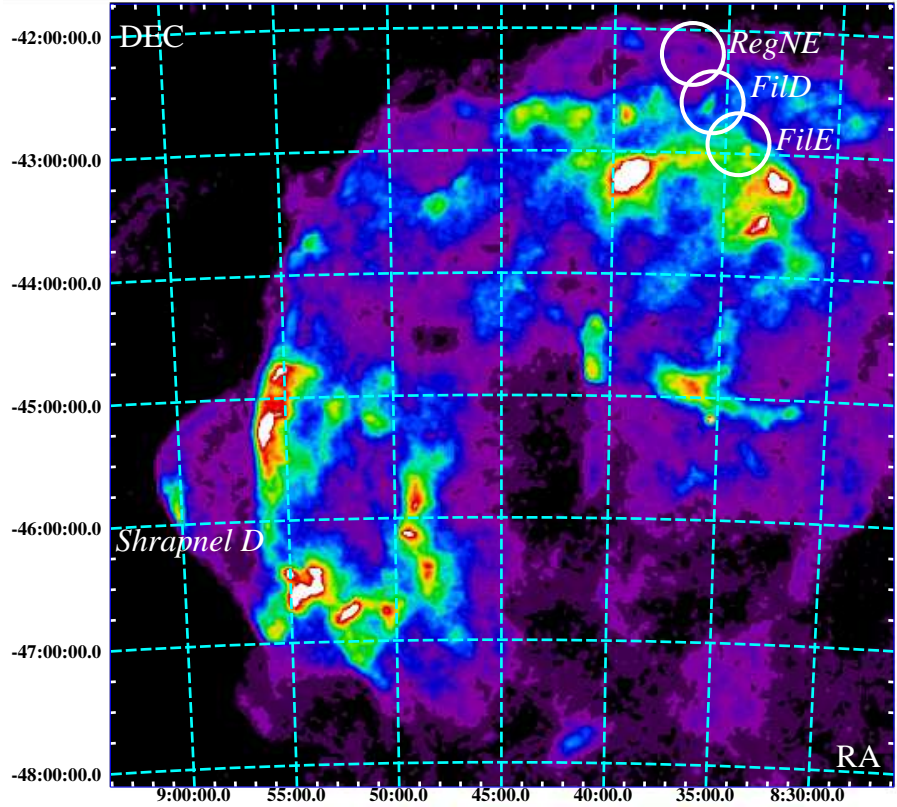}}
\caption{$ROSAT$ All-Sky-Survey image of the Vela SNR in the $0.1-2.4$ keV band. The color scale ranges between 0 and 33 counts, the bin-size is $45''$ and the image has been smoothed by a gaussian distribution with $\sigma=67.5''$. The red circles indicate the \emph{XMM-Newton} EPIC fields of view of the three observations (pointing the regions named $RegNE$, $FilD$, and $FilE$) discussed in this paper. The position of Shrapnel D is also indicated.}
\label{fig:vela}
\end{figure}

In this perspective we present the analysis of two new \emph{XMM-Newton} EPIC observations of the northern rim of the Vela SNR.
The observations (named $RegNE$ and $FilE$ in Fig. \ref{fig:vela}), together with the $FilD$ pointing, cover an area $\ga 10$ pc$^{2}$ (at 250 pc) and allow us to study at high resolution the spatial distribution of the physical and chemical properties of the plasma on this large scale. $RegNE$ was partially visible in the $FilD$ pointing analyzed in Paper I as a previously undetected (it was not visible in the \emph{ROSAT} PSPC observation discussed in \citealt{bms99}) bright X-ray knot, spectrally harder than $FilD$. 
As for $FilE$, instead, the $ROSAT$ All-Sky Survey indicates the presence of a large limb-brightened structure corresponding to a long (several pc) and bright H$\alpha$ filament at a larger projected distance from the main shock front than $FilD$ (see also \citealt{bms00}). 

The paper is organized as follows: in sect. \ref{The Data} we present the new data and the data analysis procedure; in Sect. \ref{The data analysis} we present the X-ray results in terms of image analysis (Sections \ref{The region morphology}$-$\ref{Equivalent width images}) and spatially resolved spectral analysis (Sect. \ref{spatially resolved spectral analysis}). Finally, Sect. \ref{Discussion} discusses the results, while our conclusions are summarized in Sect. \ref{Summary and conclusions}.

\section{Data processing}
\label{The Data}

The \emph{XMM-Newton} EPIC observations presented here consist of the $FilD$ observation presented in Paper I and of two new observations: i) Observation ID 0203960101 (PI F. Bocchino), with pointing coordinate $\alpha$ $(2000)=8^{h}36^{m}45^{s}$ and $\delta$
$(2000)=-42^\circ11'39''$ (i. e. the $RegNE$ region), and ii) Observation ID 0302190101 (PI M. Miceli), with pointing coordinate $\alpha$ $(2000)=8^{h}34^{m}15^{s}$ and $\delta$ $(2000)=-42^\circ55'40''$ (i. e. the $FilE$ region). The observations have been performed with the EPIC MOS (\citealt{taa01}) cameras and with the EPIC pn (\citealt{sbd01}) camera on 2004 October 31 (0203960101) and on 2006 November 14 (0302190101).  The location of the two pointings is indicated in Fig. \ref{fig:vela}, while the relevant information about the two data sets are summarized in Table \ref{tab:NR data}.

The data were processed using the Science Analysis System (SAS V7.0). Light curves, images, and spectra, were created by selecting events with PATTERN$\le$12 for the MOS cameras, PATTERN$\le$4 for the pn camera, and FLAG=0 for both. To eliminate the contamination by soft proton flares, the data were screened by applying a count-rate limit on the light curves (binned at 100 s) at high energies ($10-12$ keV for MOS and $12-14$ keV for the pn camera). The count-rate limit is 0.18 counts per second for the MOS cameras, and 0.3 counts per second for the pn camera. The screened exposure times are given in Table \ref{tab:NR data}.  

All the images presented here are superpositions (obtained using the $EMOSAIC$ task) of the MOS1, MOS2, and pn images of the two pointings and of the $FilD$ pointing presented in Paper I and are background-subtracted, vignetting-corrected, and adaptively smoothed (with the task $ASMOOTH$). 
The contribution of the background was derived from the high signal-to-noise background event files E1$\_$fm0000$\_$M1.fits, E1$\_$00fm00$\_$M2.fits, E1$\_$0000fm$\_$PN.fits (obtained using the medium filter) described in detail by \citet{rp03}.

The exposure and vignetting corrections were performed by dividing the count images by the corresponding superposed exposure maps (obtained with the task $EEXPMAP$). Since the ratio between the MOS and pn effective area $A_{rel}$ depends on the spectral distribution of the incoming photons, we have taken the thermal spectrum of the source into account when scaling the pn exposure maps by $A_{rel}$, to make MOS-equivalent superposed count-rate images. In particular, since in Paper I we found that in the $FilD$ region the emission below 0.5 keV is associated with the soft component (at $\sim 10^{6}$ K), while above $0.5$ keV the hotter component (at $\sim 3\times10^{6}$ K) dominates, we assume an incoming thermal spectrum at $10^{6}$ K for the images in the $0.3-0.5$ keV band ($A_{rel}=0.16$) and at $3\times10^{6}$ K for the images in the $0.5-1$ keV ($A_{rel}=0.20$), $0.85-0.98$ keV ($A_{rel}=0.26$), and $1-1.24$ keV ($A_{rel}=0.31$) bands. The background- and continuum-subtracted line images and the equivalent width maps presented in Sect. \ref{Equivalent width images} were produced by following the procedure described in \citet{mdb06}.
 
Spectral analysis was performed in the energy band $0.3-1.5$ keV using XSPEC. The Ancillary Response Files were generated with the SAS $ARFGEN$ task, and the event files were processed using the $EVIGWEIGHT$ task (\citealt{ana01}) to correct vignetting effects. The background contribution was subtracted from the same region positions on the CCD (i. e. in the detector coordinates), in order to take into account the inhomogeneous response of the instruments across the field of view. Spectra were rebinned to achieve a signal-to-noise ratio per bin $>5\sigma$ and the fittings were performed simultaneously on both MOS spectra and on the pn spectrum. Since the $FilE$ observation was performed after the micrometeoroid damage to the MOS1 camera, we used only the pn and MOS2 spectra in region 6 of Sect. \ref{spatially resolved spectral analysis}, because this region partially covers the damaged chip 6. We also added a systematic 5\% error term to reflect the estimated uncertainties in the calibration of the instrumental effective area (\citealt{kir07}). All the reported errors are at 90\% confidence, according to \citet{lmb76}.

\section{The data analysis}
\label{The data analysis}

\subsection{The region morphology}
\label{The region morphology}
\begin{figure*}[htb!]
 \centerline{\hbox{     
     \psfig{figure=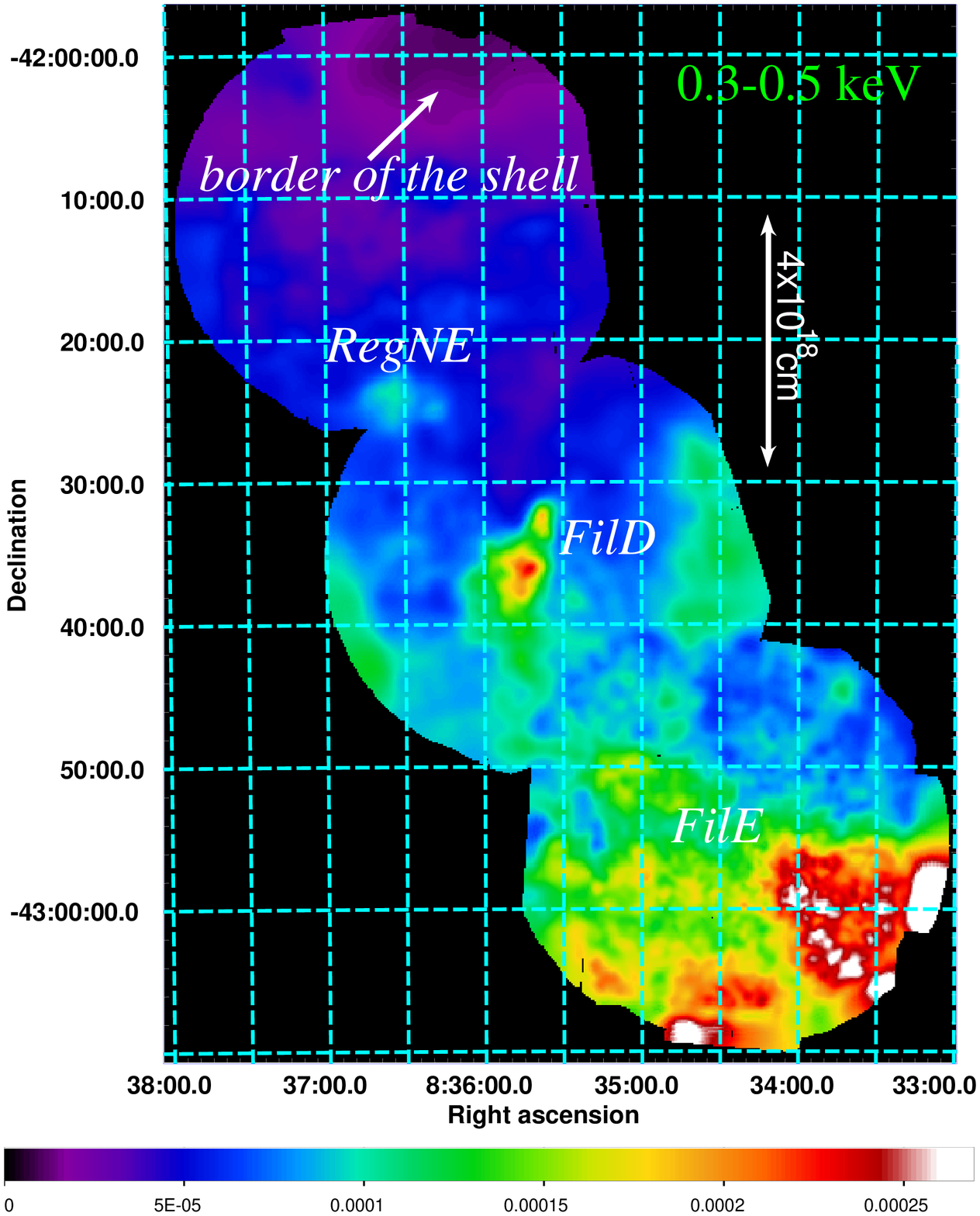,width=8.2cm}
     \psfig{figure=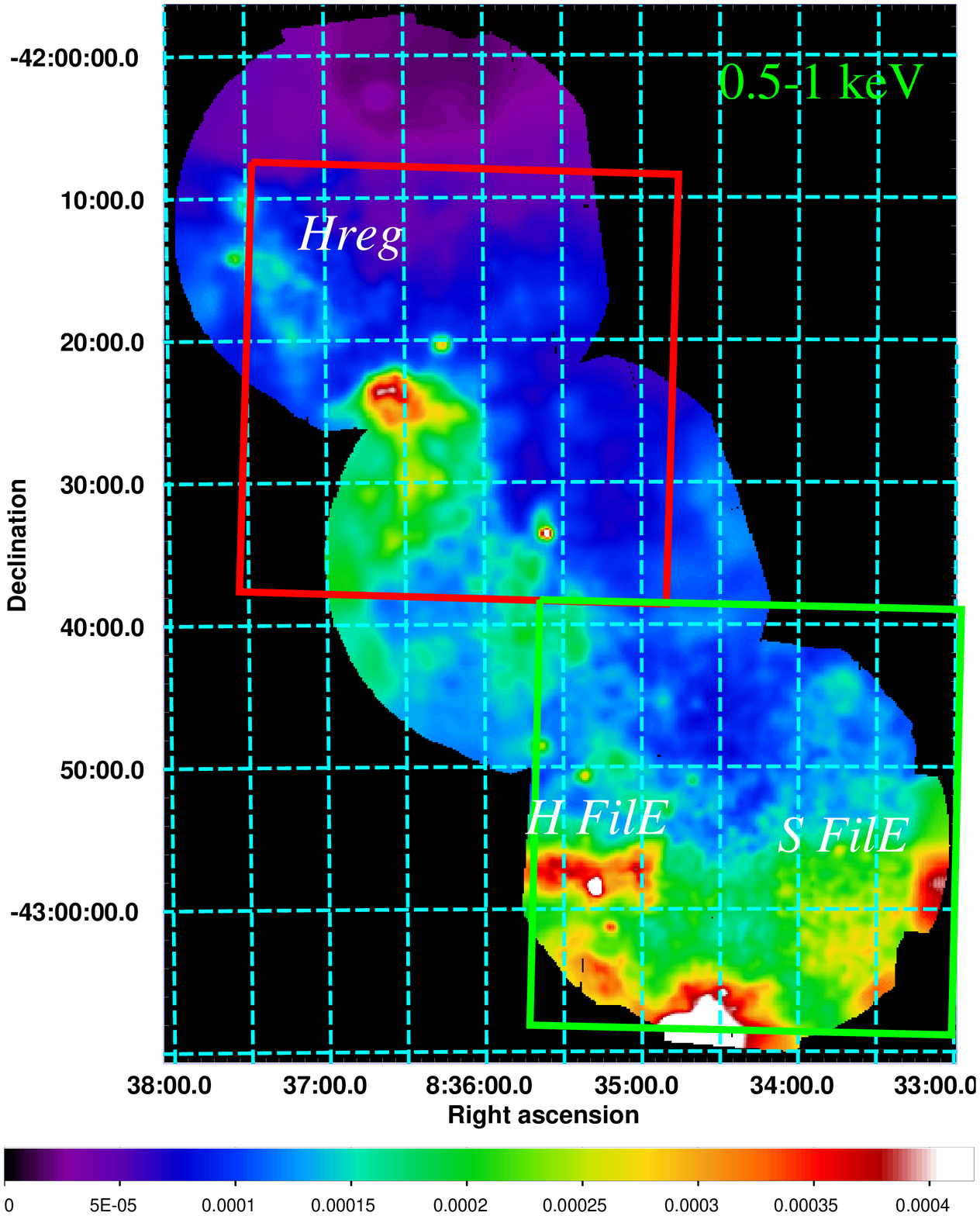,width=8.2cm}     
 }}
\caption{\emph{Left panel}: mosaiced count-rate images (MOS-equivalent counts per second per bin) in the $0.3-0.5$ keV band (bin size$=8''$) of the northern rim of the vela SNR. The image is adaptively smoothed to a signal-to-noise ratio 30. The $4\times10^{18}$ cm scale has been obtained assuming a distance $D=250$ pc. \emph{Right panel}: same as left panel in the $0.5-1$ keV energy band. 
The red and green boxes bound the field of view of the upper and lower panels of Fig. \ref{fig:NR optRegNEFilE}, respectively.}
\label{fig:NR 03051}
\end{figure*}

Figure \ref{fig:NR 03051} shows the mosaiced count-rate image of the three observations of the northern rim of the Vela shell in the two bands chosen for the study of $FilD$ in Paper I: $0.3-0.5$ keV and $0.5-1$ keV (hereafter soft band and hard band).

The bulk of the X-ray emission of the $RegNE$ knot is above $0.5$ keV, although its brightest part, at the northern end of the knot, is also visible in the $0.3-0.5$ keV band. This bright ``head'' shows a bow-shaped morphology in the $0.5-1$ keV band and lies outside the field of view of the $FilD$ observation discussed in Paper I. South of this head there is an elongated, sharp ``tail'', extending for about $15'$ (corresponding to $\sim 1$ pc, at 250 pc) towards the center of the remnant. North of $RegNE$ a different faint, diffuse region is visible in the $0.5-1$ keV band. This region, labelled $Hreg$ in Fig. \ref{fig:NR 03051}, does not present significant emission in the soft band.

In the $FilE$ region we instead observe significant emission both below and above 0.5 keV. In particular, in the Eastern side of the field of view there is a region (labelled \emph{H FilE} in Fig. \ref{fig:NR 03051}), where the surface brightness in the $0.5-1$ keV band is, on average, twice larger than that in the $0.3-0.5$ keV band. The X-ray emission of this region is harder than that in the western part of the field of view (region \emph{S FilE} in Fig. \ref{fig:NR 03051}), where we observe a high surface brightness in the soft energy band.

\begin{figure}[htb!]
 \centerline{\hbox{     
     \psfig{figure=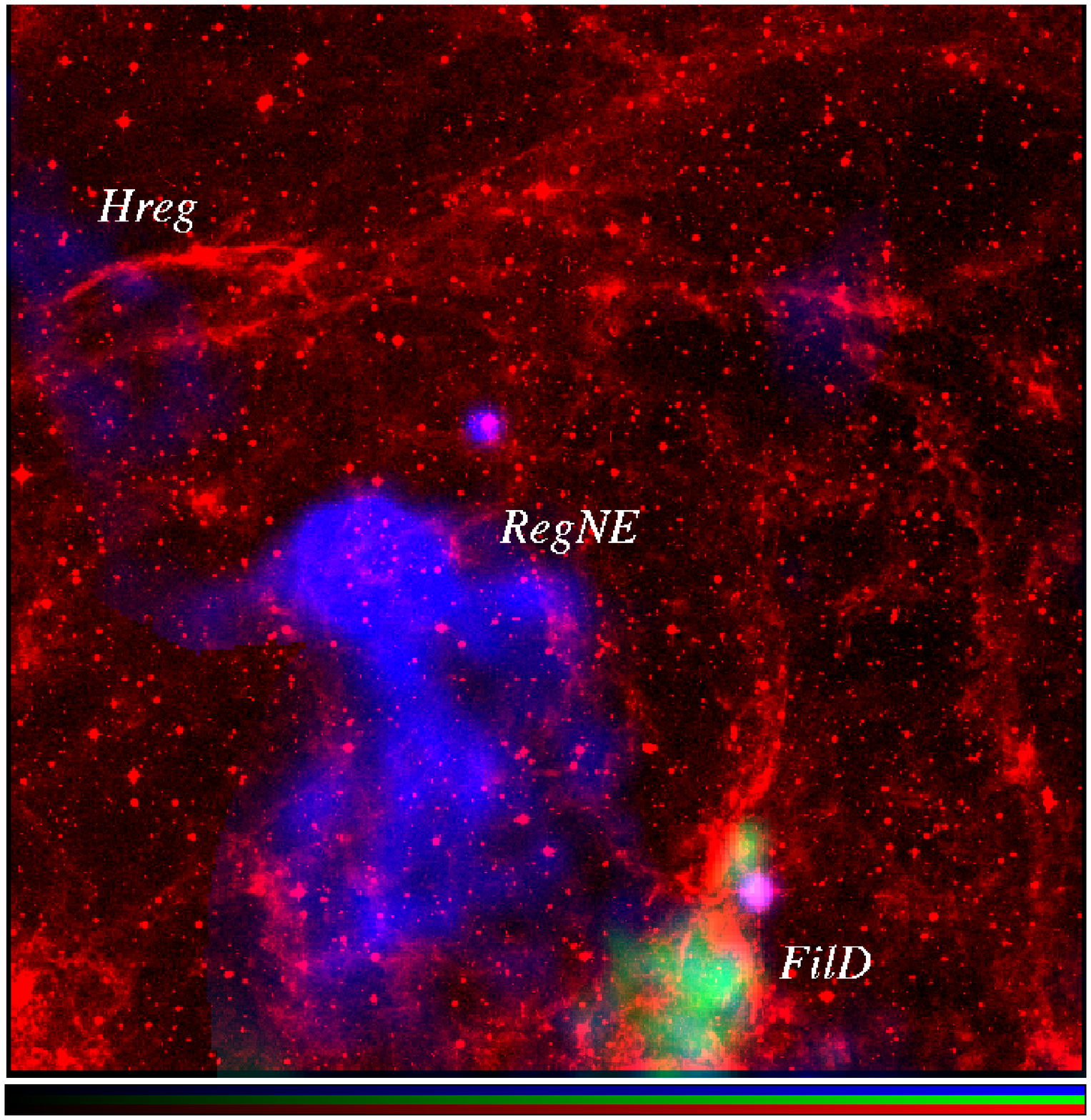,width=8.2cm}}}
      \centerline{\hbox{     
     \psfig{figure=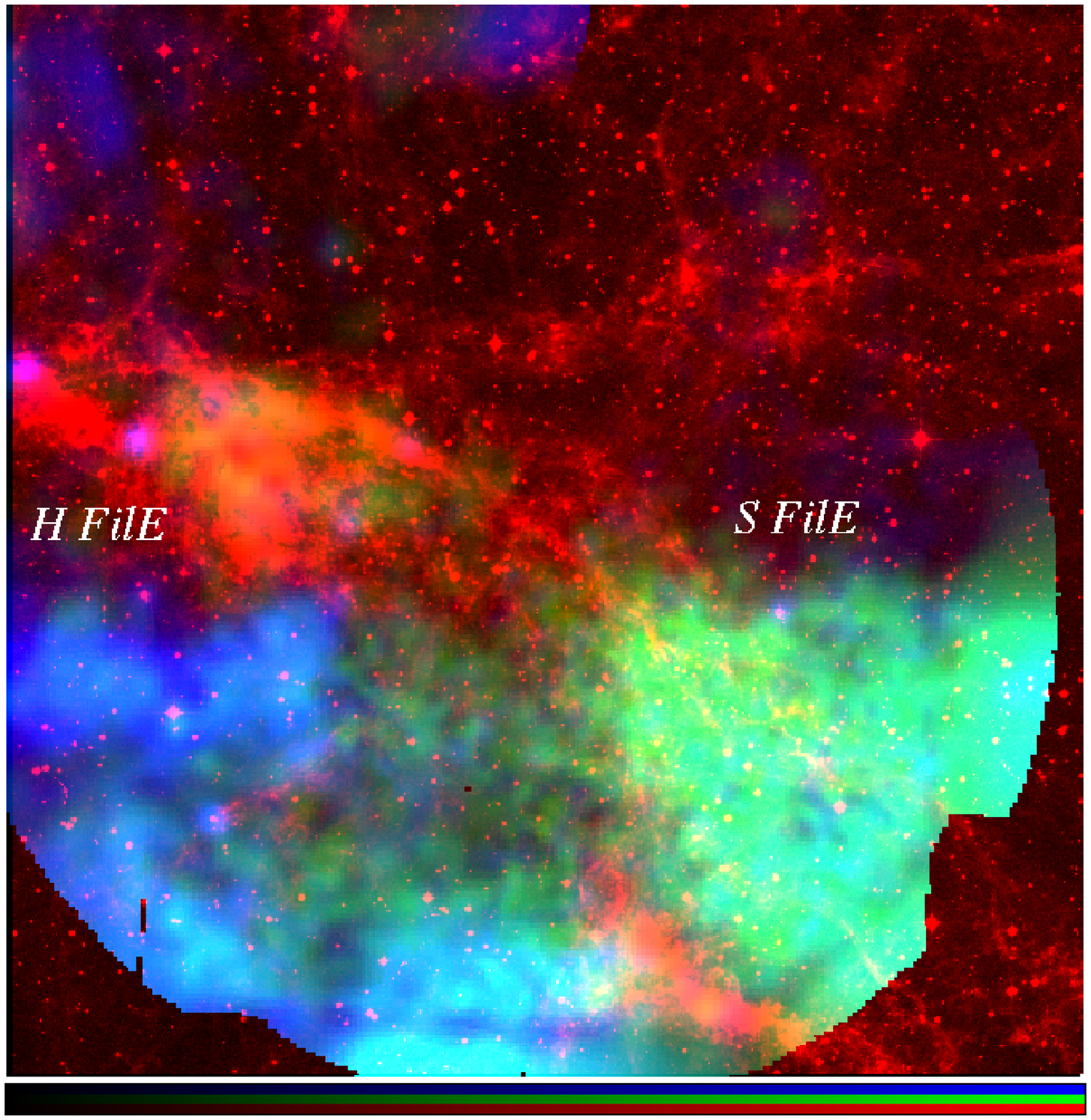,width=8.2cm}
    }}
\caption{\emph{Upper panel}: combined optical and X-ray image of $RegNE$. The H$\alpha$ emission (from AAO/UKST H-alpha survey) is indicated in red. We have superimposed, in green, the X-ray emission in the $0.3-0.5$ keV band (Fig. \ref{fig:NR 03051}, left panel) and, in blue, the X-ray emission in the $0.5-1$ keV band (Fig. \ref{fig:NR 03051}, right panel). The color bars have linear scales and range between 25\% and 60\% of the maximum count-rate for the X-ray maps, while the optical image is scaled according to the IRAF zscale algorithm. The field of view of this image is indicated by the red box in Fig. \ref{fig:NR 03051}.
\emph{Lower panel}: same as upper panel for $FilE$. The field of view of this image is indicated by the green box in Fig. \ref{fig:NR 03051}.}
\label{fig:NR optRegNEFilE}
\end{figure}

Figure \ref{fig:NR optRegNEFilE} shows a combined optical (H$\alpha$, in red) and X-ray images of $RegNE$ and $FilE$ in the soft (in green) and hard (in blue) X-ray bands. In $RegNE$ (upper panel of Fig. \ref{fig:NR optRegNEFilE}) no significant optical emission is present in the wide region corresponding to the hard X-ray emitting knot, at odd with the $FilD$ knot where both soft X-ray and optical emission are present. A couple of optical filaments seem instead related to $Hreg$.
The lower panel of Fig. \ref{fig:NR optRegNEFilE} shows the intense H${\alpha}$ emission of the $FilE$ region (in red). The \emph{H FilE} structure with hard X-ray emission, lies clearly ``behind'' the optical filament (in agreement with the results obtained by \citealt{bms00} in the same region). This configuration seems to indicate that we are observing in X-rays the plasma behind a reflected shock front. According to this interpretation, the reflected shock was generated by the impact of the main shock with a dense cloud, where the slow transmitted shock originates the optical emission. Note, however, that (as shown in Fig. \ref{fig:NR 03051}) the angular distance of this region from the border of the shell is quite large ($\ga 50'$, corresponding to $\ga 3.6$ pc at 250 pc) and that projection effects may be present. Therefore we cannot rule out the possibility that the optical filament E is not physically associated with the X-ray emitting knot.
The stratification between optical and X-ray emission is not visible in the \emph{S FilE} region, where we observe both optical and soft X-ray emission, with narrow optical filaments that lie not behind, but ``inside'' the X-ray emitting region.

\subsection{Analysis of the photon energy map}
\label{Analysis of the photon energy map}
\begin{figure}[htb!]
 \centerline{\hbox{     
     \psfig{figure=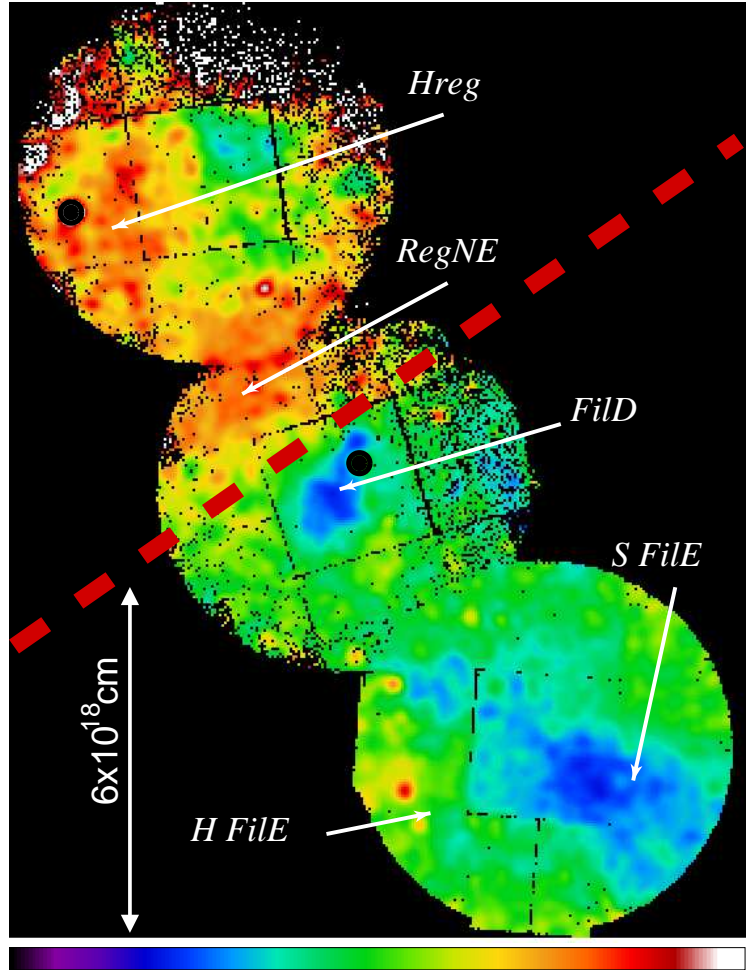,width=10cm}
 }}
\caption{MOS median photon energy map in the $0.3-2$ keV band (bin size$=12''$). In each pixel the local median photon energy of the photons detected by the two MOS cameras is reported. The pixels where the median is computed with less than 5 photons have been masked out. The color bar has a linear scale and ranges between 415 eV and 750 eV. The red dashed line indicates the border between the two regimes discussed in Sect. \ref{Analysis of the photon energy map}. Two point-like, bright X-ray sources have been masked out with black circles in the upper two fields of view.}
\label{fig:NR avgE}
\end{figure}

To obtain information about the thermal structure of the plasma, we produce the MOS median photon energy map (MPE map, i.e. an image where each pixel holds the median energy of the detected MOS photons) in the $0.3-2$ keV band. The MPE map can be considered as an indicator of the hardness of the spectrum and of the average temperature of the plasma, but it also depends on the chemical composition of the plasma and on the local value of $N_{\rm H}$ (where the interstellar absorption is higher we observe a hardening of the spectra). Note however, that, as shown in Sect. \ref{spatially resolved spectral analysis}, $N_{\rm H}$ is quite uniform in the region of the Vela shell covered by our set of observations, (in agreement with \citealt{la00}).
The median energy map is shown in Fig. \ref{fig:NR avgE} and has a bin size $12''$, so we can examine the different structures with high spatial resolution. The map clearly shows that very soft X-ray emission originates in the $FilD$ and \emph{S FilE} regions, where we observe relatively low values of MPE ($\la 500$ eV). The \emph{H FilE} region seems instead to be, on average, hotter than \emph{S FilE} and $FilD$.
We observe the maxima of the median photon energy in $RegNE$ and $Hreg$, (i. e. the regions where the bulk of X-ray emission is in the hard band). We then expect the plasma to be, on average, hotter in these two regions than in all the others.
We can also focus on the overall trend of the median photon energy on a large scale. Figure \ref{fig:NR avgE} clearly shows that two well-separate ``regimes'' (the red dashed line marks the separation in the figure) are present: a hard regime at North and a softer regime from $FilD$ to the South. 

\subsection{Equivalent width images}
\label{Equivalent width images}

To study the spatial distribution of the chemical composition of the X-ray emitting plasma, we produce equivalent width (EW) maps. These depend linearly on the abundances, but are also influenced by the temperature, column density, and ionization age. However, since temperature, $N_H$, and ionization age can be regarded as uniform (see Sect. \ref{spatially resolved spectral analysis}), the EW maps can be considered as reliable indicators of the metal abundances. As shown in Paper I, O, Ne, and Fe lines mainly contribute to the observed X-ray emission in this region of the shell, but thanks to the good statistics of the new observations we now detect also the presence of Mg lines. More specifically, in all spectra the K-shell line complexes of O VII (at $0.56$ keV), O VIII (at $0.65$ keV), Ne IX (at $0.92$ keV), and Mg XI (at $1.35$ keV) are clearly visible. We here focus only on the Ne IX (i. e. $0.85-0.98$ keV energy band) and Mg XI ($1.29-1.45$ keV energy band) emission lines, since it is possible to evaluate the continuum in an adjacent band ($1-1.24$ keV) and since the equivalent width of these lines will be less affected by local variations of $N_{H}$ than the low-energy O lines. The EW images were constructed
by dividing the background- and continuum-subtracted line images by
the corresponding underlying continuum. To estimate the continuum under the lines, we scale the continuum band adjacent to the line emission by using as a phenomenological model a thermal bremsstrahlung at $kT=0.2$ keV (absorbed by a column density $2.3\times10^{20}$ cm$^{-2}$) for the continuum, plus narrow Gaussian components for the lines. We verified that this ad-hoc model provides a satisfying description for the spectra in all the regions presented in Sect. \ref{spatially resolved spectral analysis} (reduced $\chi^2\sim 1.1-1.8$, with $\sim150-200$ d. o. f. ). A word of caution should be spent for the Ne IX EW, since this emission line emerges over a ``false-continuum'' generated by the blending of the Fe XVII L-lines (which have energies in the range $\sim 0.72-1.1$ keV). The Ne IX EW will not therefore depend simply on the Ne abundance, but on the ratio between the Ne and the Fe abundances. Note also that in the continuum energy band ($1-1.24$ keV) the emission may be contaminated by Fe and/or Ni L lines and this may yield to an over-estimation of the continuum and then to an under-estimation of our EWs.

Figure \ref{fig:EQW} shows the Ne IX and Mg XI equivalent width maps. In both maps huge inhomogeneities in the equivalent width are visible. At odd with the median photon energy map, we do not observe a North-South anisotropy, while an East-West anisotropy is present. In particular, three knots (corresponding to $Hreg$, $RegNE$, and partially \emph{H FilE}) with enhanced values of EW are present in both maps. Since the EW depends linearly on the abundance, the presence of inhomogeneities in the equivalent width maps suggests that the X-ray emission in this part of the Vela shell cannot be associated with a chemically uniform medium.

\begin{figure}[h!]
 \centerline{\hbox{	
     \psfig{figure=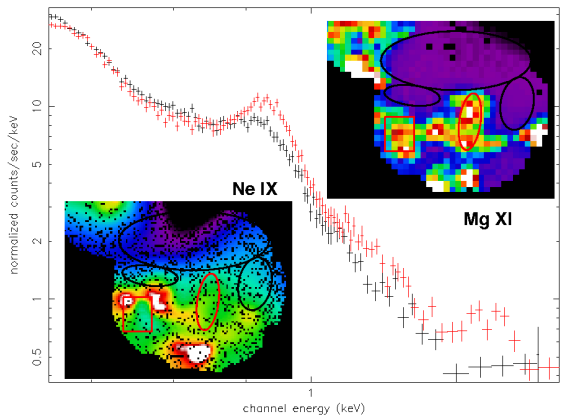,width=8.2cm}}
     }
\caption{pn spectrum (in red) extracted from regions with high Ne IX and Mg XI equivalent width (red regions shown in the EW maps in the insets) compared with the pn spectrum (in black) extracted from regions with low EW (black regions). This shows the differences between metal rich regions and ISM regions.}
\label{fig:test}
\end{figure}

Figure \ref{fig:test} includes the spectra extracted from two regions with high and low Ne IX and Mg XI equivalent width. The areas of the two regions have been chosen so as to have almost the same counts in the continuum band. Figure \ref{fig:test} clearly shows that the Ne IX and Mg XI emission lines are locally highly enhanced.

\subsection{Spatially resolved spectral analysis}
\label{spatially resolved spectral analysis}

\begin{figure}[h!]
 \centerline{\hbox{     
     \psfig{figure=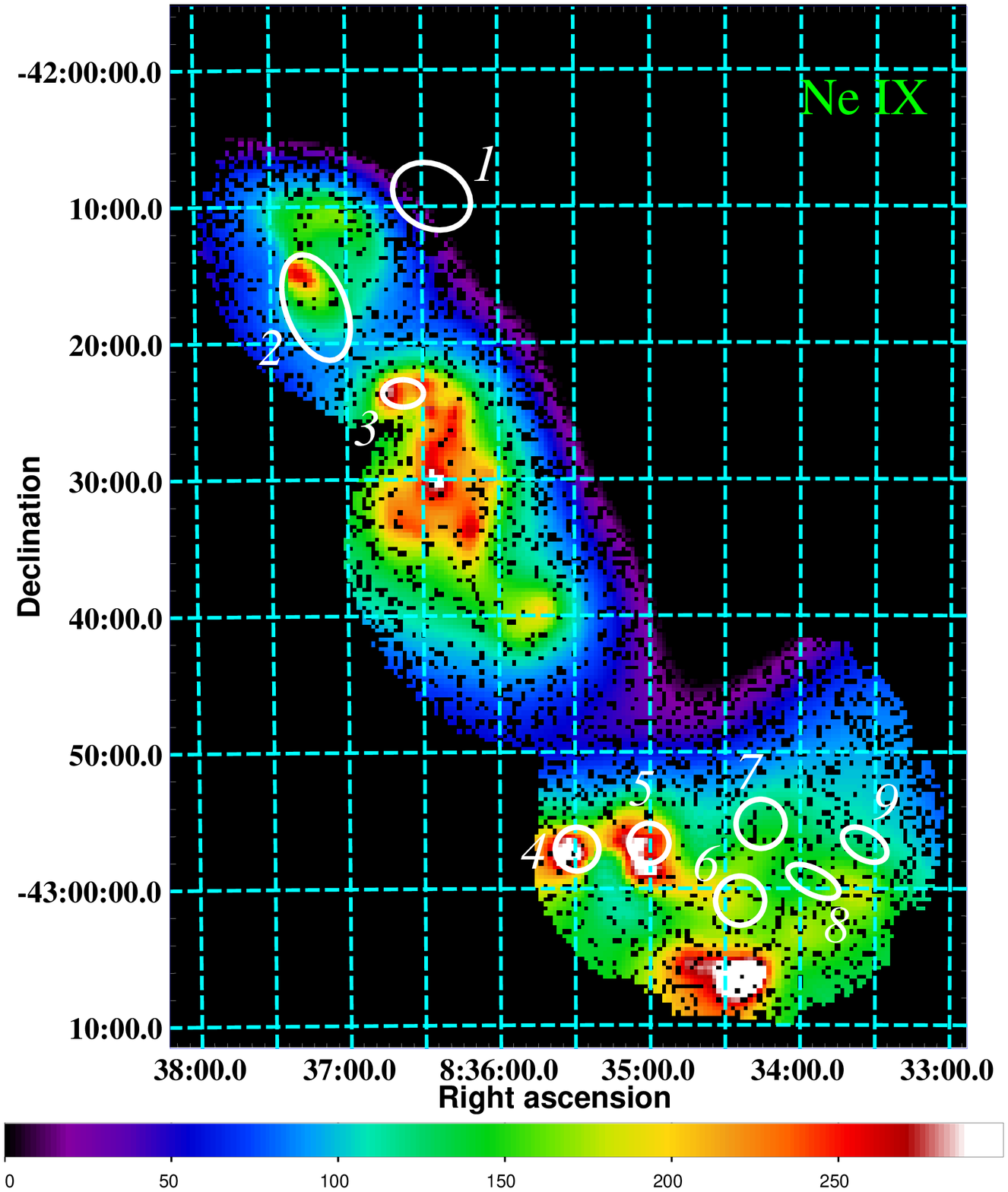,width=8.2cm}}}
      \centerline{\hbox{     
     \psfig{figure=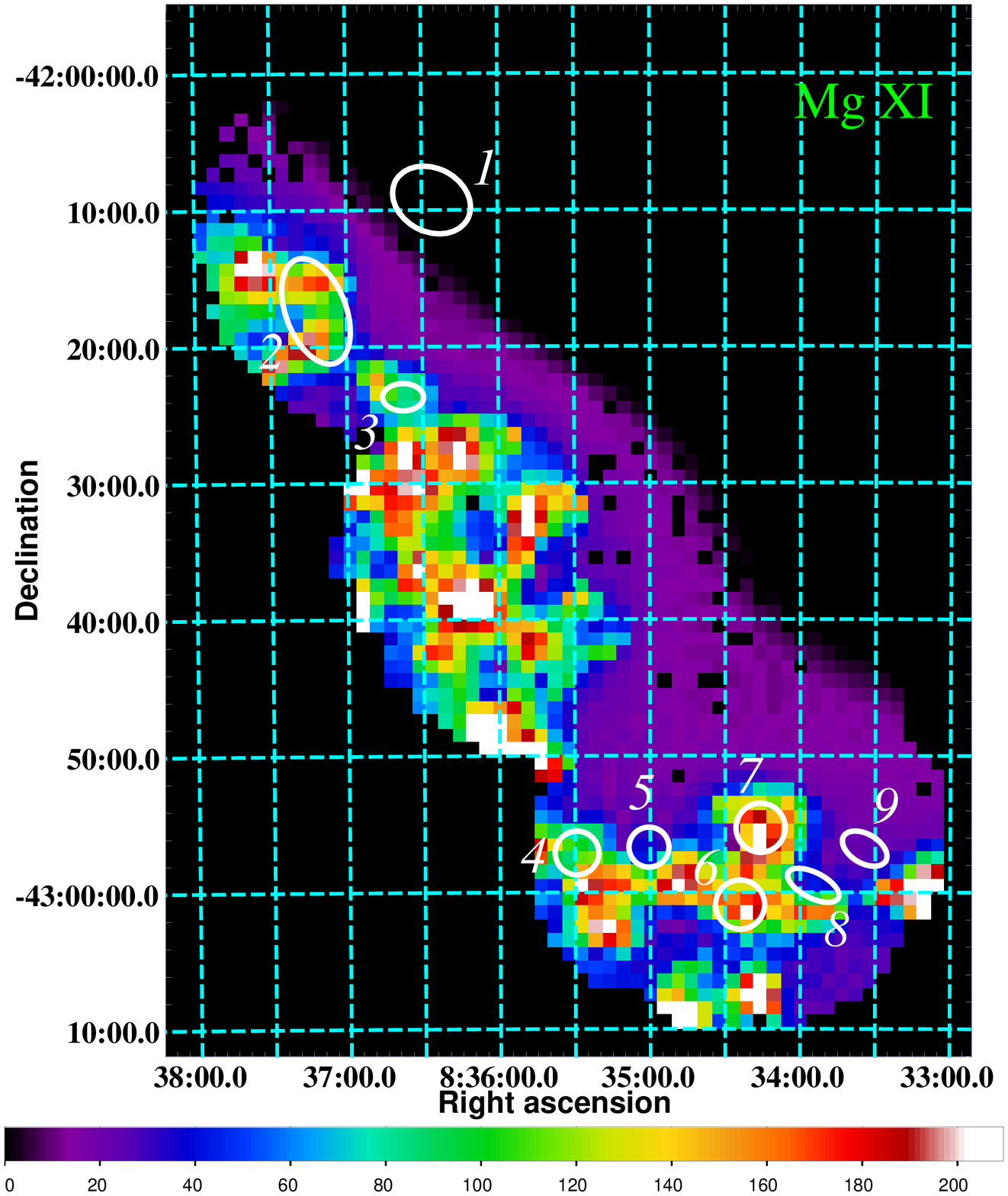,width=8.2cm}
    }}
\caption{\emph{Upper panel:} equivalent width maps of the Ne IX line emission in the $0.85-0.98$ keV band. The bin-size is $20''$ and the equivalent width ranges between 0 and 300 eV. The points where the continuum-subtracted line emission is consistent with the background have been masked out. The 9 regions selected for spatially resolved spectral analysis are also shown. \emph{Lower panel:} same as upper panel for the Mg XI line emission in the $1.29-1.45$ keV band, the bin-size is $1'$ and the equivalent width ranges between 0 and 210 eV.}
\label{fig:EQW}
\end{figure}

To check whether the spatial variations in the EWs are due to inhomogeneities in the chemical composition of the X-ray emitting plasma or to other effects (e. g. variations in temperature and/or in $N_{H}$), we perform a spatially resolved spectral analysis. 
As explained in Paper I, the poor statistics of the $FilD$ observation did not allow us to perform an accurate study of the metal abundances and we aim at exploiting the higher statistics of the new observations to address this issue.
We analyze the spectra extracted from the 9 regions shown in Fig.~\ref{fig:EQW}. The shape and size of the spectral regions were chosen in order to extract the spectra from chemically and physically uniform regions. To this end, we select regions with fairly uniform values of the Ne IX and Mg XI equivalent width (as shown in Fig. \ref{fig:EQW}) and, at the same time, very small fluctuations of the median photon energy ($\le 3\%$).

Region 1 corresponds to the low-surface-brightness region immediately behind the border of the shell, region 2 to the hard X-ray emitting $Hreg$ knot, and region 3 corresponds to the brightest part (the ``head'') of $RegNE$. Regions 4 and 5 are located in the \emph{H FilE} structure, regions 8 and 9 in the bright, soft X-ray emitting \emph{S FilE} knot, while regions 6 and 7 are located between the two parts of $FilE$, where we observe large values of the Mg XI equivalent width. We observe large EW both for Ne IX and Mg XI in regions 2, 3, and 4. In region 5 we have large Ne IX EW and low Mg XI EW, while we observe relatively low values both for neon and magnesium in regions 1, 8, and 9.

\begin{figure}[htb!]
 \centerline{\hbox{     
     \psfig{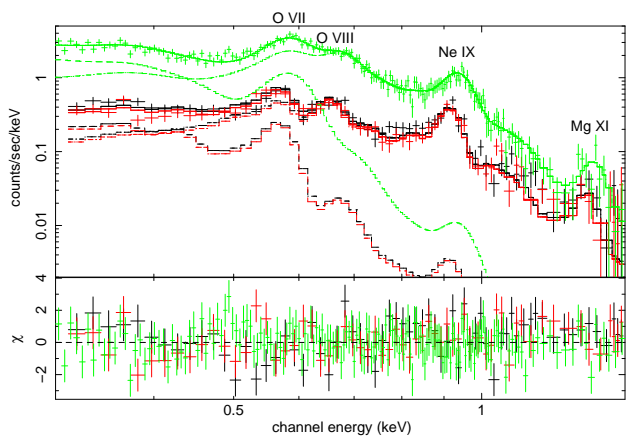}}}
\caption{pn spectrum (upper, in green) and MOS1,2 (lower, in black and red) spectra of region 4 of Fig. \ref{fig:EQW} with the corresponding best-fit model and residuals obtained with two thermal components. The contribution of the two components and the ions that mainly contribute to the line emission are indicated.}
\label{fig:spec4}
\end{figure}

All the extracted spectra are described well by two MEKAL components of an optically-thin thermal plasma in collisional ionization equilibrium (\citealt{mgv85}, \citealt{mlv86}, \citealt{log95}) and the temperatures of the two components are quite similar to those found in Paper I for the $FilD$ knot (i. e. $T_{I}\sim 10^6$ K and $T_{II}\sim 3\times10^{6}$ K, respectively). Figure \ref{fig:spec4} shows a representative spectrum (extracted from region 4 of Fig. \ref{fig:EQW}) with its best-fit model and residual. As shown in Fig. \ref{fig:spec4}, strong O VIII, Ne IX, Mg XI, and Fe XVII line (which have energies in the range $\sim 0.72-1.1$ keV) complexes are associated with the hotter component. Therefore, we leave the O, Ne, Mg, and Fe abundances free in the hotter component. Since the line emission mainly originates in the hotter component, it is not possible to determine the chemical abundances in the cold component. Therefore the abundances of the cooler component are frozen to the solar values (\citealt{ag89}). This assumption is in agreement with the results obtained in Paper II, where we showed that in the FilD knot the cold component is associated with interstellar cloud material heated by the transmitted shock front.
Except for region 3, all spectra are described significantly better by this model than by a single temperature model (either in CIE or in non-equilibrium of ionization). 

In region 3 (i. e. the head of $RegNE$) one component is enough. In Paper I we showed that also the spectra extracted from the tail of $RegNE$ (regions 14 and 15 in Paper I) are well described by a single thermal component in CIE. However, we found (Paper I, Sect. 4.6) that also a non-equilibrium of ionization (NEI) model at higher temperature ($\sim 7 \times10^{6}$ K) can be appropriate to describe the spectra. We try to overcome this ambiguity by modeling the pn and MOS spectra of the head of $RegNE$ (where the statistics are significantly better than in the tail) with a single MEKAL model and with the XSPEC NEI model. We do not obtain significant improvement in the quality of the fits by using the NEI model ($\chi^{2}_{CIE}=264.8$, with 243 d. o. f. and $\chi^{2}_{NEI}=264.2$, with 242 d. o. f.). We then adopt the CIE spectral model. 

We find an entanglement between the best-fit values of $N_H$ and the temperature of the cooler component, $T_I$, with large error bars for both the parameters. To overcome this problem we determine the $N_H$ value in region 3 (where only the hotter component is present), thus finding $N_{H}^{reg3}= 2.3^{+0.8}_{-1}\times10^{20}$ cm$^{-2}$. We then fix $N_H$ to this value for the fittings of the spectra extracted from the other regions. We checked that the assumption of a uniform $N_H$ in the whole observed part of the shell is valid. In fact, the $\chi^{2}$ minima obtained by letting the $N_H$ parameter free are not significantly lower (according to the F-test) than those obtained by fixing it to the best-fit value obtained in region 3\footnote{For example, in region 1, the best-fit value of $N_H$ is consistent with $N_{H}^{reg3}$ and we obtain very similar vales of $\chi^2$ by letting the $N_H$ parameter free ($\chi^{2}_{free}=222.4$, with 206 d. o. f.) and by fixing it to $N_H^{reg3}$ ($\chi^{2}_{frozen}=222.5$, with 207 d. o. f.). Even in region 8, where we obtain a very low best-fit value of $N_H$ (consistent with 0 and lower than $1\times10^{19}$ cm$^{-2}$, with $\chi^{2}_{free}=259.9$ and 212 d. o. f.), the quality of the fit does not change significantly by putting $N_H=N_H^{reg3}$ ($\chi^{2}_{frozen}=263.0$, with 213 d. o. f.).}. 

The best-fit parameters for all 9 spectral regions are given in Table~\ref{tab:bestfit}.
The temperatures of the two X-ray emitting components are quite similar to those found in the $FilD$ region of Paper I, with nearly uniform values both for $T_{I}\sim0.9-1.2\times10^{6}$ K and $T_{II}\sim2-3\times 10^{6}$ K.
We instead observe large variations (more than one order of magnitude) in the emission measures. This result confirms that, as shown in Paper I and Paper II, neither component can originate in the intercloud medium. The emission measure per unit area ($EM$) of the cold component shows low values ($\la 10^{18}$ cm$^{-5}$) at North, in the $RegNE$ pointing, and large values (many $10^{18}$ cm$^{-5}$) in the $FilE$ region. The ratio $EM_I/EM_{II}$ ranges between $EM_I/EM_{II}\sim2.5$, in the regions with the hardest X-ray emission (i. e. region 2, or $Hreg$ in Fig. \ref{fig:NR 03051}, and regions 4 and 5, or $HFilE$ in Fig. \ref{fig:NR 03051}) and $EM_I/EM_{II}\ga 5$ in the softest regions (regions 7 and 8, i. e. the brightest part of $SFilE$ in Fig. \ref{fig:NR 03051}). These results allow us to understand the double-regime in the MPE map: the region North of $FilD$ presents larger values of the median photon energy because in that region the cold component is fractionally small. 
 
The results of the spatially resolved spectral analysis (Table~\ref{tab:bestfit}) further confirm that \emph{the X-ray emitting medium is not chemically homogeneous}. We observe large inhomogeneities in the O, Ne, Mg, and Fe abundances, with significant overabundances both for Ne and Mg. We remark that the metal abundances do not change significantly if we adopt the NEI model. As explained in Sect. \ref{Equivalent width images}, we expect the Ne IX equivalent width map to depend on the Ne$/$Fe abundances. We indeed find a very good agreement between the Ne IX equivalent width map shown Fig. \ref{fig:EQW} and the best-fit Ne$/$Fe abundances shown in Table~\ref{tab:bestfit}. For example, we find the highest best-fit values of Ne$/$Fe (Ne$/$Fe$>5$) in regions 3, 4, and 5, where the Ne IX EW map has three maxima, while we find Ne$/$Fe$\la2$ in regions 1 and 9, where the Ne IX equivalent width is lower. The agreement between the Mg abundance and the Mg XI EW map of Fig. \ref{fig:EQW} is very good. We find significantly enhanced Mg abundances in regions 2, 3, 4, 6, while the Mg abundances is consistent with the solar value where the Mg XI equivalent width is lower (regions 1, 5, 8, 9, see Fig. \ref{fig:EQW} and Table~\ref{tab:bestfit}).

\section{Discussion}
\label{Discussion}
The combined analysis of the three \emph{XMM-Newton} observations of the northern rim of the Vela shell reveals the presence of physical and chemical inhomogeneities in the X-ray emitting medium.

Although all spectra are described well by two thermal components at $\sim10^6$ K and $\sim 3\times10^6$ K, a wide region at $\la 6\times 10^{18}$ cm from the border of the shell presents a higher median photon energy than that observed at larger distances from the main shock front. This is due to the lower values of the emission measure of the cold component and indicates that in this region the plasma is, on average, hotter and, if we assume pressure equilibrium, also more rarefied. At larger distances we instead observe bright optical filaments and large, dense, X-ray emitting clouds (with larger values of the emission measure of the cold component), together with a soft diffuse and faint emission maybe associated with small, unresolved, and dense cloudlets.

Both the EW maps and the spatially resolved spectral analysis unequivocally show that the chemical composition of the X-ray emitting plasma is not homogeneous. These results clearly rule out the possibility that the whole X-ray emission originates in the shocked interstellar medium. Moreover the spatially resolved spectral analysis confirms the detection of several knots with overabundant Ne and Mg. 
This strongly suggests we are indeed observing ejecta. It would also be possible that we were observing clumps of wind residuals from the progenitor star shocked by the main shock front. However, stellar winds from massive stars typically show overabundances of light elements, like He and N (\citealt{ev92} and \citealt{cgg03}) and it would be difficult to explain overabundances of Ne and Mg according to this scenario. We then suggest we are indeed observing SN ejecta.
This indication is further supported by the fact that the regions with high EWs in Fig. \ref{fig:EQW} (and high Ne and Mg abundances) are visually aligned toward the center of the shell. 

The detection of X-ray emitting ejecta associated with the evolved Vela SNR is not totally unexpected. \citet{aet95} have observed 6 shrapnels of X-ray emitting material outside the Vela shell (Shrapnels A-F). Two of these have been investigated in detail (Shrapnel D, \citealt{kt05} and Shrapnel A, \citealt{kt06}). In Shrapnel D large overabundances of O (O/O$_\odot \sim6.4$), Ne (Ne/Ne$_\odot \sim13.7$), and Mg (Mg/Mg$_\odot \sim14$) have been found by \citet{kt05}, who also revealed a slight Fe overabundance (Fe/Fe$_\odot \sim1.4$). 
As shown in \citet{kt06}, Shrapnel A is instead very rich in Si (Si/Si$_\odot \sim2.6$), while O, Ne, Mg, and Fe have solar or subsolar abundances. Note that the angular distance from the center of the remnant to Shrapnel A is bigger by $\sim20\%$ than that to Shrapnel D. The abundances in Shrapnel A indicate an efficient mixing between the fast-moving shrapnel and the swept-up interstellar medium. 
In both cases, however, the chemical composition suggests that the shrapnels are fast ejecta.

We remark that in our case we would be observing the first shrapnels ever detected $within$ the Vela shell (Shrapnels A-F are all outside the border of the remnant).
Note, however, that although the projected position of the ejecta is behind the main shock front, it is clearly possible for these ``new'' shrapnels to be physically outside the border of the shell. Only shrapnels ejected approximately on the plane of the sky have been detected so far, but it is very likely that many other shrapnels have been ejected also in other directions. We then expect to find fragments of ejecta in other regions of the remnant. When shrapnels are observed in projection within the shell, the X-ray emission is not fully associated with the ejecta, because (since the plasma is optically thin) there is also the contribution of the shocked ISM intercepted along the line of sight. Therefore, the enhanced Ne and Mg abundances we find in regions 2, 3, 4, 6, and 7 are indeed weighted averages of the ejecta and ISM abundances. Nevertheless, we can compare our abundances with those found in Shrapnel D (Fig. \ref{fig:vela}) by \citet{kt05}. In particular, we consider the relative abundances between heavy elements, which are less dependent on the spectral model than the absolute abundances. In regions 2 ($Hreg$) and especially in region 3 ($RegNE$) the relative abundances (normalized to the Fe abundance) are quite similar to those observed in Shrapnel D: O:Ne:Mg:Fe~$\sim$~2.0:5.0:6.3:1 (region 2) and O:Ne:Mg:Fe~$\sim$~2.0:6.7:9.0:1 (region 3), to be compared with O:Ne:Mg:Fe~$\sim$~4.6:9.8:10.0:1 (Shrapnel D, \citealt{kt05}). This result supports the association of $Hreg$ and $RegNE$ with ``new'' shrapnels. Although in region 4 the relative abundances do not decrease significantly (O:Ne:Mg:Fe~$\sim$~1.5:5.4:5.8:1), 
at South (specially in regions 6 and 7), where we expect to intercept more shocked ISM along the line of sight, the relative abundances become lower (O:Ne:Mg:Fe~$\sim$~0.7:2.2:3.5:1 in region 6; O:Ne:Mg:Fe~$\sim$~0.8:2.1:4.7:1 in region 7). 
Note that the largest deviations between our shrapnels and Shrapnel D occur in the O/Fe abundances. The cause may be that since the X-ray emission in the Vela SNR is quite soft, we expect the contribution of the post-shock ISM to be stronger in the low-energy line complexes (i. e. O VII and O VIII) than in the Ne IX and Mg XI lines. 
As for the Si-rich Shrapnel A, \citet{kt06} found subsolar or solar values for O, Ne, Mg, and Fe, with lower values of the abundance ratios (O:Ne:Mg:Fe~$\sim$~0.4:1.0:0.8:1). Although we cannot estimate the Si abundances in our observations, we can conclude that our new shrapnels are intrinsically different from Shrapnel A.

In the regions where the emission seems to be completely associated with the interstellar medium (regions 8 and 9) we find indications for a metal depletion in the X-ray emitting plasma, since the O, Ne, and Fe abundances are significantly lower than solar (see Table \ref{tab:bestfit}).

\section{Summary and conclusions}
\label{Summary and conclusions}

The combined analysis of the three \emph{XMM-Newton} observations of the northern part of the Vela shell allowed us to obtain a high-resolution description of a large area ($\ga 10$ pc$^{2}$) of the remnant. 

We observe hard X-ray emitting knots at North, while there are large, soft X-ray emitting knots associated with bright optical filaments closer to the center of the shell.

All spectra are described well with two thermal components whose temperatures are almost uniform in the whole region. 
The observed inhomogeneities of the emission measures confirm that, as in the case of $FilD$, the two components cannot be associated with the uniform intercloud medium.

We find inhomogeneous patterns of abundances of O, Ne, Mg, and Fe and we detect different knots ($RegNE$, $Hreg$ and part of $FilE$) with significant overabundances of Ne and Mg. The observed pattern of abundances can hardly be reconnected with the only interstellar medium and/or clumps of wind residuals.
We conclude that these knots are previously undetected shrapnels of ejecta which appear, in projection, to be inside the Vela shell. 
We expect the X-ray emission from these new shrapnels to be ``contaminated'' by the X-ray emission from the shocked ISM intercepted along the line of sight. Nevertheless, in the regions where we expect this effect to be less considerable ($RegNE$ and $Hreg$), the relative abundances  are quite similar to those found in Shrapnel D by \citet{kt05}.

In conclusion, we confirm the presence of two phases in the X-ray emitting plasma (at $\sim10^6$ K and $\la 3\times10^6$ K), as found in Paper I. In addition, we argue that the metal abundances in the hot component are not uniformly distributed. Regions with oversolar Ne and Mg are reasonably associated with shrapnels of ejecta similar to Shrapnel D. On the other hand, the other regions are probably associated with the evaporation of shocked interstellar clouds, according to the scenario described in Paper II.

\acknowledgements
We thank the referee for his helpful comments and suggestions.
This work was partially supported by Agenzia Spaziale Italiana, grant number ASI/INAF I/023/05/0.
This work makes use of results produced by the PI2S2 Project managed by the Consorzio COMETA, a project co-funded by the Italian Ministry of University and Research (MIUR) within the Piano Operativo Nazionale ``Ricerca Scientifica, Sviluppo Tecnologico, Alta Formazione'' (PON 2000-2006). More information is available at http://www.pi2s2.it and http://www.consorzio-cometa.it.

\bibliographystyle{apj}

\begin{deluxetable}{lccccc}
\tablecaption{Relevant information about the data.}
\tablewidth{0pt}
\tablehead{
\colhead{OBS$\_$ID} & \colhead{CAMERA} & \colhead{$t_{exp}$ (ks)$^{*}$} & \colhead{Mode}  & \colhead{Filter}
}
\startdata
0203960101 &  MOS1   &       52.1/26.3        &    Full Frame   &  medium  \\ 
0203960101 &  MOS2   &       52.1/27.3        &    Full Frame   &  medium  \\
0203960101 &   pn    &       48.6/18.2        &    Ext. Full Frame     &  medium  \\
0302190101 &  MOS1   &       39.8/32.7	      &    Full Frame   &  medium  \\ 
0302190101 &  MOS2   &       40.0/28.5        &    Full Frame   &  medium  \\
0302190101 &   pn    &       35.8/23.7        &    Ext. Full Frame	  &  medium  \\
\enddata
\tablenotetext{*}{Unscreened/Screened exposure time.}
\label{tab:NR data}
\end{deluxetable}

\clearpage
\thispagestyle{empty}
\begin{deluxetable}{lcccccccccc} 
\rotate
\tablecaption{Best-fit parameters for the spectra extracted in the 9 regions shown in Fig. \ref{fig:EQW}, described with two thermal components in collisional ionization equilibrium (only one component in region 3). The $N_H$ parameter is fixed to $2.3\times10^{20}$ cm$^{-2}$ (as explained in Sect. \ref{spatially resolved spectral analysis}). All errors are at the 90\% confidence level.}
\tablewidth{0pt}
\tablehead{
 \colhead{Parameter}  &  \colhead{Region 1} & \colhead{Region 2}  & \colhead{Region 3}   & \colhead{Region 4}  &  \colhead{Region 5} &  \colhead{Region 6}  & \colhead{Region 7} &  \colhead{Region 8} & \colhead{Region 9}
}
\startdata 
$T_{I}$ ($10^6$ K)& $1.06^{+0.03}_{-0.02}$ & $1.12\pm0.07$ & $-$ & $1.05^{+0.04}_{-0.03}$&$1.07^{+0.03}_{-0.05}$ & $0.95^{+0.04}_{-0.05}$ & $0.97^{+0.04}_{-0.02}$ & $1.06\pm0.02$&$1.03^{+0.05}_{-0.04}$\\ 
$EM^{*}_{I}$  & $1.4^{+0.2}_{-0.4}$ & $0.77\pm0.08$ & $-$ & $4.1^{+1}_{-0.3}$ & $3.9^{+0.5}_{-0.3}$ & $9.9^{+0.6}_{-0.5}$ & $6.5^{+0.6}_{-1}$ & $8.3\pm0.2$ & $5.5^{+0.3}_{-0.4}$\\
$T_{II}$ ($10^6$ K)&$2.1^{+0.3}_{-0.2}$&$2.94^{+0.04}_{-0.1}$&$2.58\pm0.08$&$2.62^{+0.04}_{-0.1}$&$2.64^{+0.05}_{-0.1}$&$2.12^{+0.07}_{-0.06}$&$2.29^{+0.1}_{-0.06}$&$2.56^{+0.05}_{-0.04}$&$2.35^{+0.2}_{-0.09}$\\
$EM^{*}_{II}$ & $0.20^{+0.13}_{-0.09}$ & $0.33\pm0.02$ & $1.3^{+0.7}_{-0.2}$ & $1.45^{+0.09}_{-0.05}$ & $1.64^{+0.15}_{-0.07}$ & $3.4^{+0.7}_{-0.2}$ &$1.3^{+0.1}_{-0.2}$& $1.48^{+0.07}_{-0.12}$&$2.4^{+0.2}_{-0.4}$\\
$O/O_\odot$ &$0.74^{+0.06}_{-0.08}$&$1.1\pm0.1$&$1.0\pm0.1$&$0.74^{+0.03}_{-0.05}$&$0.57^{+0.04}_{-0.08}$&$0.40^{+0.01}_{-0.04}$&$0.35^{+0.02}_{-0.03}$&$0.25\pm0.02$&$0.25\pm0.02$\\
$Ne/Ne_\odot$  & $2.6^{+0.3}_{-0.4}$ & $2.8\pm0.3$& $3.3^{+0.4}_{-0.8}$& $2.7^{+0.2}_{-0.1}$& $1.7^{+0.3}_{-0.1}$& $1.3\pm0.1$ & $0.9^{+0.1}_{-0.2}$ &  $0.5\pm0.1$  & $0.5\pm0.1$ \\
$Mg/Mg_\odot$     &   $2\pm2$    & $3.5^{+0.5}_{-0.9}$ & $4.4\pm1.3$ &    $2.9\pm0.8$   &  $1.3^{+0.6}_{-0.5}$  &  $2.1^{+0.9}_{-0.7}$ &  $2.0^{+0.4}_{-0.7}$  & $0.7\pm0.4$ & $0.5^{+0.5}_{-0.4}$\\
$Fe/Fe_\odot$ &$1.4^{+0.5}_{-0.2}$&$0.56^{+0.1}_{-0.05}$&$0.49^{+0.9}_{-0.6}$&$0.50^{+0.05}_{-0.04}$&$0.32^{+0.08}_{-0.05}$&$0.6^{+0.2}_{-0.1}$&$0.43^{+0.07}_{-0.08}$&$0.21^{+0.03}_{-0.02}$&$0.25^{+0.03}_{-0.07}$\\    
$\chi^{2}/$d. o. f.& $222.5/207$ &   $309.9/306$   &  $264.8/243$   &   $296.2/291$    &   $337.2/263$   &   $174.9/193$   &    $309.0/234$    &  $263.0/213$    &  $281.4/218   $\\    
\enddata
\tablenotetext{*}{Emission measure per unit area in units of $10^{18}$ cm$^{-5}$.}
\label{tab:bestfit}
\end{deluxetable}

\clearpage

\end{document}